\documentstyle[pre,aps,preprint]{revtex}
\begin{document}

\title{Dynamical properties of the hypercell spin glass model}
\author{P. M. Gleiser F. A. Tamarit 
\thanks{E-mail:  pgleiser@fis.uncor.edu and tamarit@fis.uncor.edu}} 
\address{Facultad de Matem\'atica, Astronom\'{\i}a y F\'{\i}sica \\
        Universidad Nacional de C\'ordoba \\
        Ciudad Universitaria, 5000 C\'ordoba, Argentina}
\date{\today}
\maketitle
\begin{abstract}
The spreading of damage technique is used to study 
the sensibility to initial conditions in a heath
bath Monte Carlo simulation of the spin glass hypercubic cell model.
Since the hypercubic cell in dimension 2D and the hypercubic lattice in
dimension D resemble each other closely at finite dimensions and both
converge to mean field when dimension goes to infinity, it allows us
to study the effect of dimensionality on the dynamical behavior of
spin glasses.
\end{abstract} 


%\renewcommand{\arraystretch}{0.6}

%\newpage
\section{Introduction}
\label{introduction}
The spin glass theory has been one of the most difficult problems
treated by statistical mechanics during the last two decades. 
Despite its value in the field of solid state physics, its
study has also contributed to develop new techniques which now apply
to a wide range of fields, such as optimization problems, neural
networks, and other complex systems \cite{Fischer}.

The first microscopic approach to spin glasses is due to Edwards and
Anderson (EA) \cite{Edwards}, whose model basically consists on a Ising
system with random positive and negative exchange couplings.  Until
now only its mean field version, known as the Sherrington Kirkpatrick
model (SK) \cite{Sherrington} has been exactly solved, but
unfortunatly its solution requires the sophisticated replica trick. 
Under such limitations, Monte Carlo
numerical simulations have become one of the most applied techniques
in the field.  At the same time, it is not clear at the moment
whether one should expect that the spin glass phase of the mean field
SK model resembles the behavior of the spin
glass phase of the low dimension EA model.

For many years there has been great controversy on whether the spin
glass transition is either of thermodynamical or dynamical nature.
However numerical simulations \cite{Ogie} and phenomenological 
scaling arguments at zero temperature \cite{Bray} strongly suggest the
existence of a true thermodynamical phase transition. From a
dynamical point of view, a very careful numerical study of the time
decay of the auto--correlation function $q(t)$ has shown that the
system displays three different dynamical regimes:
above the Curie point $T_c$ of the nonrandom Ising model, the
auto-correlation decays exponentially;
between $T_c$ and the spin glass temperature $T_g$ the
auto-correlation function has a stretched exponential behavior
with temperature dependent exponents;
finally, in the spin glass phase only power law decay is
observed at all times scales.
 
Since the SK model can be understood as the infinite dimension 
version of the EA model, it is desirable to be able to study the
effects of dimensionality both in the static and dynamical properties
of the system, even if such analysis should be limited to numerical
considerations. In 1992  Parisi, Ritort and Rub\'{\i} \cite{pariru}
introduced the {\em hypercubic cell model} which allows a very
efficient treatment of high dimensional models, at least when
compared with hypercubic lattice models. It consists of
a unique cubic cell of dimension $D$ with an Isin spin variable
associated to each of its $2^D$ cornes. Despite its simplicity and
unrealistic features, one expects that its behavior for
dimension $2D$, resembles, at least qualitatively, that observed in
a $D$ dimensional hypercubic lattice, since both share the same
connectivity $D$. Even more, for $D\to \infty$ the hypercell model
recovers the mean field SK model. 

This approach has been used in the last few years to analyze both
dynamical and statical consequences of dimensionality in the spin
glass phase of different models
\cite{pariru,Cugliandolo,Marinari,Stariolo}. In this work we apply
the damage spreading method to the hypercell Ising spin glass model
simulated with a heat bath Monte Carlo dynamics. This technique
basically consists in measuring the time evolution of the Hamming
distance between two initially different configurations submitted to
the same thermal noise, i.e, updated with the same random number
sequence. The dependence of the damage and other related quantities
on temperature, time, initial conditions and other relevant
parameters leads to a dynamical phase diagram of the model.

In general, this phase diagram strongly depends on the Monte Carlo
dynamics used in the numerical simulation. In particular, for the
bi and three-dimensional Ising ferromagnet one finds that the dynamical
transition coincides with the static one when the system is submitted to
heat  bath dynamics, while the opposite occurs when submitted to Glauber
dynamics. When more complex systems are analyzed with heat bath
dynamics, more than two dynamical phases are usually found, where 
only a few of them are correlated with termodynamical phases
(see \cite{Silva} and references therein). In particular, for spin
glasses in three 
and four dimensions \cite{derwei}, three different dynamical regimes were
obtained, as ocurred when the auto-correlation was analyzed. 
For low temperatures, the final damage is non null and its value
depends on the initial Hamming distance. For intermediate
temperatures, the damage still spreads but its final value is
independent on the initial damage. Finally, for high temperature the
final damage is always zero.  While the lower dynamical transition
temperature seems to agree with the equilibrium one ($T_g$) separating the
spin glass and the paramagnetic phases, the upper transition
temperature seems to be consistent with the temperature below which
the stretched exponential relaxation emerges ($T_c$). Surprisingly, a similar
behavior was reported for the two dimensional spin glass model, which
does not present a non zero temperature spin glass phase. Nevertheless,
despites some numerical evidence, it is not clear  whether
these three regimes found with damage spreading are related to those
oberved through the temporal behavior of the auto--correlation 
function. On the other hand, when the SK model was studied 
only a two phases structure was found, in good agreement with the
thermodynamical diagram. 

This paper is organized as follows. In section 2 we introduce in more
detail the hypercell model and describe the spreading of damage
technique. In section 3 we present the results for different
dimensions. Finally, in section 4 we discuss the main conclusions of
the paper. 

\section{The model and the method}
\label{model}
The model consists of a single hypercubic cell in dimension $D$ with
an Ising spin variable $S_i = \pm 1$ associated to each of its $2^D$
corners. Each spin interacts with its $D$ nearest neighbours through
the Hamiltonian
\begin{equation}
H= - \sum_{\langle ij \rangle} J_{ij} S_i S_j ,
\end{equation}
where $\langle ij \rangle$ denotes nearest neighbours and the
$J_{ij}$ are chosen accordingly to the following probability
distribution:
\begin{equation}
P_J(J_{ij})=\frac{1}{2}\delta(J_{ij}-J)+\frac{1}{2}\delta(J_{ij}+J). 
\end{equation}
Here we have taken $J=1/D^{1/2}$ to normalize extensive quantities. \\

The method consists on simulating the time evolution of the system through
a heath bath Monte Carlo process. The spins are sequentially updated with 
the following rule:
\begin{equation}
S_i(t+1)= 
\left\{
\begin{array}{ll}
+1 & \qquad \mbox{with probability} \quad \frac{1}{2}[1+\tanh(h_i(t))] \\
 & \\
-1 & \qquad \mbox{with probability} \quad \frac{1}{2}[1-\tanh(h_i(t))] \\
\end{array}
\right.
\end{equation}
where $h_i= \sum_{i \ne j}^N J_{ij} S_i(t)$ is the local field in
site $i$ at time $t$.

For a given disorder configuration $\{ J_{ij} \}$ we choose two
different initial states $\{ S_i^A \}$ and $\{ S_i^B \}$ and let both
evolve with the same thermal noise, i.e., by using the same random
sequence. We then measure the time  evolution of the Hamming distance
or {\em damage} between them, defined as
\begin{equation}
dh(t)=\frac{1}{4N} \sum_{i}^N (S_i^A(t)-S_i^B(t))^2 .
\end{equation}
For each temperature we calculate the time average of the damage
$\overline{\langle dh \rangle}$ over $10 000$ Monte Carlo Steps (MCS),
defined by:
\begin{equation}
\overline{\langle dh \rangle} = \frac{\sum_{t} \langle dh
\rangle(t) 
P(t)}{\sum_{t} P(t) } \; .
\end{equation}
Here $P(t)$ is the probability that the two replicas do not become
identical at time $t$ \cite{derwei}.  Note that, since we use the same
random sequence for updating both replicas, if at any time $t$ they
become identical (i.e., they meet in the phase space), they will
continue identical for all subsequent times.
 
This procedure was repeated $M$ times ($M$ depending on $D$ and $T$) in
order to obtain  a configurational average $\langle dh \rangle$ of
the damage over different coupling constants, initial conditions and
random number sequences. 

In the next section we will study the influence of the dimensionality
$D$ and the initial damage between the two replicas $dh(0)$ in the
long time behavior of the Hamming distance. This will allow us to
characterize different dynamical behaviors as a function of the
temperature of the system and analyze their possible relationships
with the thermodynamical phases.

\section{Results}
We start this section by describing the behavior of the model for
dimension $D=8$. In Fig.\ 1 we show the Hamming distance as
a function of temperature for three different initial damages, namely,
$dh(0)=0.1$, $0.5$ and $1$.  %

Observe that the system displays three
different dynamical regimes: 
\begin{description}                
\item[a)] for low temperatures ($T < T_1^8$) we observe that $\langle
dh \rangle $ is non null and its value depends on the initial damage
(it increases as the initial damage increases);
\item[b)] for intermediate temperatures ($T_{1}^{8} < T < T_{2}^{8}$)
the system is characterized by a single value of $\langle dh \rangle$
independent of the initial damage; 
\item[c)] for high temperatures ($T > T_2^8$) the  Hamming distance
$\langle dh \rangle $ is always zero.
\end{description}
This behavior is similar to the one observed by Derrida and Weisbuch
\cite{derwei} in the three-dimensional Edwards-Anderson model and
differs from that observed on the Sherrington--Kirkpatrick model, as
described in the introduction. As we will show soon, the same
qualitative behavior was also observed for $D=6$, $10$ and $15$.   

Next we characterize each phase by the temporal behavior of both
$P$ and $\langle dh \rangle$. In Fig.\ 2 we show
the behavior of $P(t)$ and $\langle dh \rangle (t)$ for $T=0.53$
(the low temperature phase), with $dh(0)=1$.

After suffering an exponential decay to a value close to $0.5$ (a
similar behavior was observed by Arcangelis {\em et al.}
\cite{arca} for the EA model) the Hamming distance $\langle dh
\rangle (t)$ grows slowly, while $P(t)$ decays slowly too. In Fig.\ 3
 we show the same results in a double logarithmic plot, from
which follows  that, after an initial transient of about 1000 MCS
both quantities vary with a power law behavior $P(t) \approx
t^{-\delta}$ (with $\delta\approx 0.258$) and $\langle dh \rangle(t) \approx
t^{-\gamma}$ (with $\gamma\approx 0.047$). We have also found that these
exponents depend on the temperature of the system, although a more
careful analysis of such dependence should be done with better
statistics and for different temperatures in order to confirm these
results.

This behavior can be understood in terms of the  phase space
structure of the system. If, as happens in the SK model, the phase
space has valleys separated by a wide distribution of Hamming
distances, then the replicas that are closer to each other become
identical faster than those that are further apart. As time goes on,
$\langle dh \rangle$ takes into account only those replicas that are far apart
and, as a consequence, it grows. This indicates that bigger energy
barriers separate valleys that are further apart. Since we are
working with small systems, $N=256$, this barriers can be crossed
for long times as the ones  we considered ($t=10000$).

Next we make a similar study in the intermediate phase. In Fig.\ 4
 we plot the curve $\ln{(-\ln{P(t)})}$ vs. $\ln{t}$ for
$T=1.41$, which, for a wide range of values of $t$, can be very well
fitted by a linear function indicating a stretched exponential decay
of $P(t)$.  The Hamming distance presents a different behavior, since
it remains constant as time flows and $P(t)$ decays. For long times
$\langle dh \rangle$ displays big fluctuations, which appear as a
consecuence of the poor statistics (note that only a small number of
replicas have survived for such long times).  These results accept
three different interpretations:
\begin{itemize}
\item the system has a phase space structure with multiple valleys but
all of them equidistant;
\item the system has only two valleys, like a ferromagnet;
\item the phase space is almost flat as a function of the free
energy, so the two replicas wander through a phase space (represented
by a hypercube of dimension $2^{2^D}$) and do not find themselves due
to its high dimensionality. 
\end{itemize}
In the first two hipotesis, the faster decay of $P(t)$ indicates that
the valleys  are not mutually impenetrable. It is
probable then that in this paramagnetic phase the system separates
regions in phase space (valleys) that are accesible to each other.

Finally, in the high temperature phase ($T > T_2^6$) all the replicas become
zero in a few MCS and both  $P(t)$ and $\langle dh \rangle$  decays
exponentially.   

This results are very important since:
\begin{itemize}
\item  $P(t)$ has a temporal behavior similar to the one found in
\cite{Ogie}, indicating the possibility of a close relationship
between the phases found with spreading of damage and those studied
through the auto--correlation function $q(t)$;
\item they show that the hipercell model in dimension $D=8$ is similar
to the three and four dimensional EA model not only in 
its static properties (as studied by Parisi {\em et al.}
\cite{pariru}) {\em but also} in its  dynamical behavior.
\end{itemize}

The same detailed study was performed for $D=6$ and the same
qualitative behavior was observed for all the quantities. 

In dimension $D=10$ a new dynamical behavior emerges.  In considering
the $\langle dh \rangle$ vs. $T$ plot presented in Fig.\ 5,
we see that the system basically displays the same three regimes
found in $D=8$. Neverthless a more detailed analysis of the
dependence of $P$ and $\langle dh \rangle$ with time, showed in Fig.\ 6, 
reveals new
features.  Now both in the intermediate and the lower temperature
phases $P(t)$ equals 1 for all considered times ($t < 10000$) while
$\langle dh \rangle$ keeps a constant value (after an initial fast
exponential decay). The only difference resides in the dependence on
the initial damage shown in Fig.\ 5.

The difference between this phases can be better observed in Fig.\ 7
 where we present the histograms of Hamming distances in
$t=100$ for $T=0.35$ and $T=1.92$ respectively, with initial damage
$dh(0)=1$. We verify that the low temperature phase still presents a
wide  distribution, indicating a complex structure as the one
described by replica symmetry breaking. On the other hand, in the
intermediate regime the distribution is narrow, indicating a behavior
that corresponds to one of the three hipotesis made for the $D=8$
case.  It is worth mentioning that the histograms present the same
qualitative behavior in all studied dimensions indicating a drastic
change in the phase space structure at the critical temperature
$T_1$.

The same analysis has been done for $D=15$ and in Fig.\ 8 we
present $\langle dh \rangle$ vs. $T$ with the three usual phases.
The temporal analysis displays the same behavior in the different
phases. 

Finally, in table \ (\ref{tabla1}) we present the values of the critical
temperatures  $T_{1}^{D}$ and $T_{2}^{D}$  obtained for the
different dimensions studied in this paper.

Note that as $D$ increases, $T_1$ seems to approach, as expected, the
value 1, which corresponds to the critical static temperature of the
Sherrington Kirkpatrick model. Unfortunatly, as much as we know, the
static critical temperatures of the spin glass--paramagnetic
transition for finite $D$ have never been studied, so, it is
impossible to compare static and dynamical transition temperatures.
If, as happens with all Ising like spin models studied
in the literature with heat bath dynamics, these
temperatures coincide, we can then conclude that the
convergence of this critical temperature $T_1$ is very slow.
Concerning $T_2$, it also increases,
but higher dimensions should be considered in order to extrapolate
the $D\to \infty$ behavior. It is important here to stress that, at
least for $D=15$, we have not found a dynamical behavior that
resembles the one obtained in the study of the Sherrington
Kirkpatric model, namely, a two phase structure with the critical
dynamical temperature in good agreement with the static one. 
In other words, for all the temperatures considered in this paper,
we have shown that the system has a dynamical phase diagram similar
to the one of the Edwards--Anderson model, i.e., we did not find an
{\em upper critical dynamical dimension} above which the system displays 
a mean field behavior.

\section{Conclusions}

In this work we have applied the damage spreading technique to the
hypercell Isin spin glass model in order to study its dynamical
behavior and the influence of dimensionality. As was stressed in the
introduction, previous studies had found different dynamical phase
diagrams for the EA and the SK model. While the former presented
three different regimes (suggesting a correlation with the temporal
decay of the autocorrelation function), the last one presented a unique
phase transition at a temperature compatible with the spin
glass--paramagnet static transition. Since the SK model is recovered
as the $D\to \infty$ version of the EA model, we studied the effect of
increasing the dimensionality in the dynamical behavior of the system
in the hope of finding some critical dimension above which the system
displays the mean field dynamical phase diagram.

The phase diagram, for all dimensions studied, presents a three phase
structure similar to that obtained for the EA model with $D=3$ and
$D=4$, namely, a low temperature phase that displays dependence with
the initial damage, an intermediate phase where the damage spreads
but its final value is independent of the initial damage and a high
temperature phase where the damage decays exponentially to zero.
While the lower critical dynamical temperature seems to converge to
SK static temperature, for the upper critical temperature we were not
able to extrapolate its behavior (we are probably far from an
asymptotic regime). This means that, at least for $D=15$, we are
still far from the SK regime. Further simulations with higher
dimensions would be required but the computation time needed exceeds
our numerical capacity.

When one considers the temporal behavior of the quantity $P(t)$ for 
different dimensions, some interesting conclusion can be extracted:
\begin{itemize}
\item There is a drastic change in the behavior of $P(t)$ for $D\le 8$ and
$D\ge 10$. In the former case, $P(t)$ displays a decay similar to
that observed for the auto-correlation function in the EA model
\cite{Ogie} characterizing three different phases: power law decay
for $T < T_{1}^{D}$, stretched exponential decay for $T_{1}^{D} < T <
T_{2}^{D}$ and exponential decay for $T > T_{2}^{D}$. In the last case
($D \ge 10$), $P(t)$ is constant and equals 1 in the low and
intermediate temperature regimes and decays exponentially in the high
temperature phase ($T > T_{2}^{D}$).
\item The detailed analysis of the histograms of Hamming distances
reveals that the low temperature phase is characterized by a wide
distribution, as expected in a multi-valley phase diagram, for all
the dimensions considered. This structure resembles, at least
qualitatively,  the one found in the SK model. On the other hand, in
the intermediate phases we always found narrow distributions of the
Hamming distances. Note that in this regime the final distance is
always non zero independently of the initial damage. This is also
true for vanishing small initial damages, meaning that in this phase
the heat bath Monte Carlo dynamics is truely chaotic.
\end{itemize}

\acknowledgements We are gratefully acknowledged to D.A. Stariolo for fruitful discussions. 

%%%%%%%%%%%%%%%%%%%%%%%%%%%%%%%%%%%%%%%%%%%%%%%%%%%%%%%%%%%%%%%%
%          Caption for figures
%%%%%%%%%%%%%%%%%%%%%%%%%%%%%%%%%%%%%%%%%%%%%%%%%%%%%%%%%%%%%%%%
%

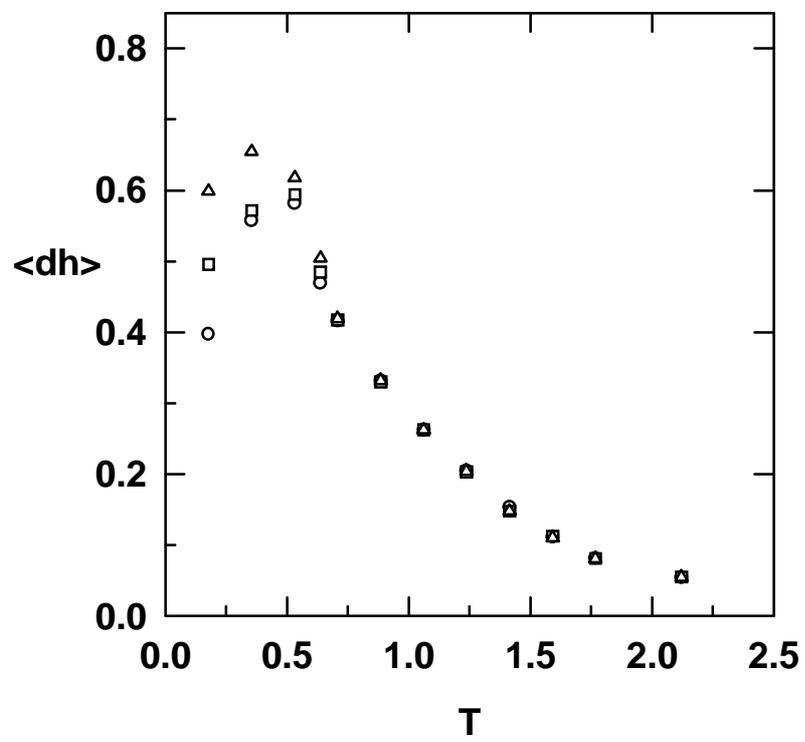
\begin{figure}
\label{fig1}
\caption{$\langle dh \rangle$ vs Temperature for $D=8$ and for three
different initial damages: $dh(0)=1$ (triangles), $dh(0)=0.5$
(squares) and $dh(0)=0.1$ (circles).}
\end{figure}

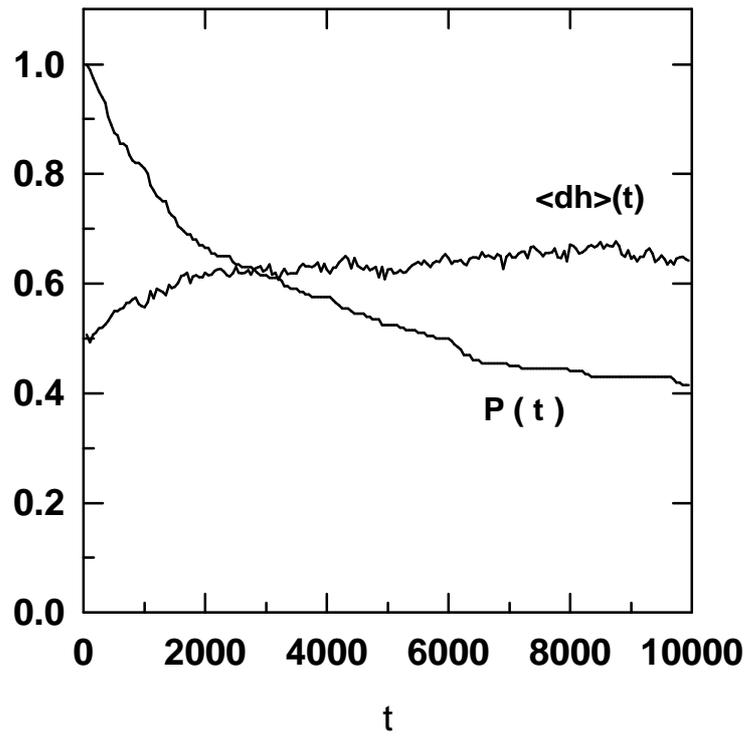
\begin{figure}
\label{fig2}
\caption{Temporal behavior of $P(t)$ and $\langle dh \rangle(t)$ for
$D=8$ and $dh(0)=1$. The average was calculated over 1000 different
samples.}
\end{figure}

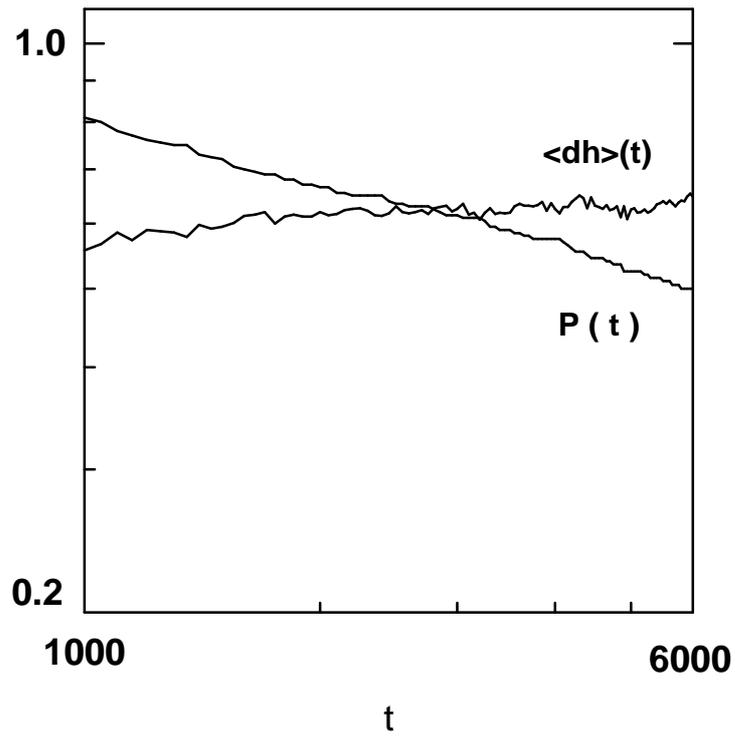
\begin{figure}
\label{fig3}
\caption{$\langle dh \rangle$ and $P(t)$ as a function of $t$ in a
double logaritmic scale for $D=8$ and $T=0.52$ (in the low
temperature phase).}
\end{figure}
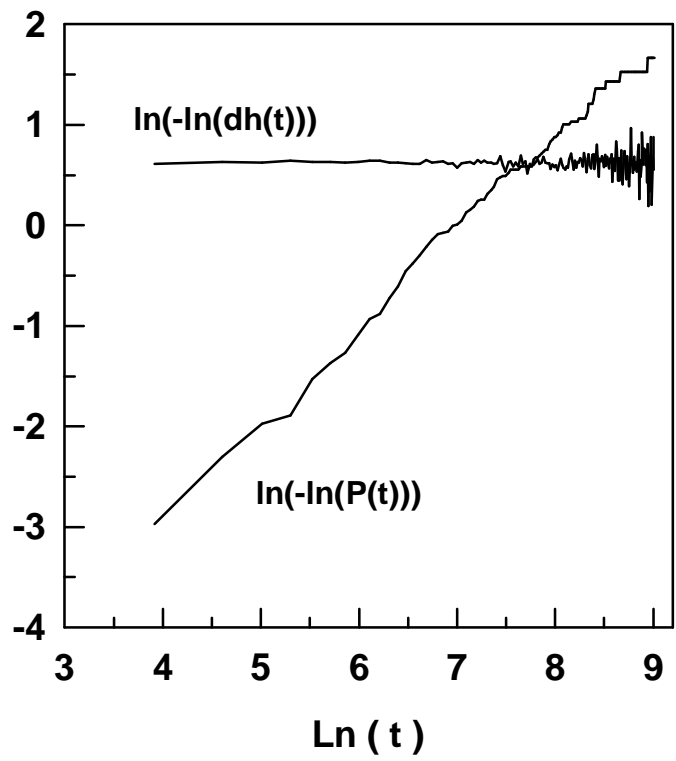
\begin{figure}
\label{fig4}
\caption{ $\ln{(-\ln{\langle dh\rangle})}$ and $\ln{(-\ln{P(t)})}$
for $D=8$ and $T=1.41$, (in the intermediate phase).}
\end{figure} 

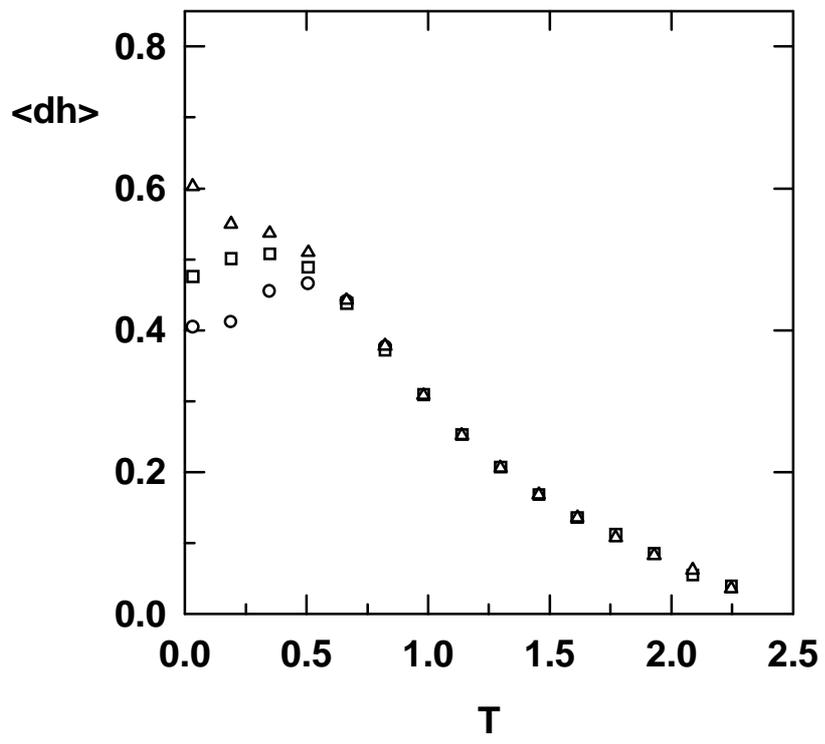
\begin{figure}
\label{fig5}
\caption{$\langle dh \rangle$ vs Temperature for $D=10$ and for three
different initial damages: $dh(0)=1$ (triangles), $dh(0)=0.5$
(squares) and $dh(0)=0.1$ (circles).}
\end{figure}

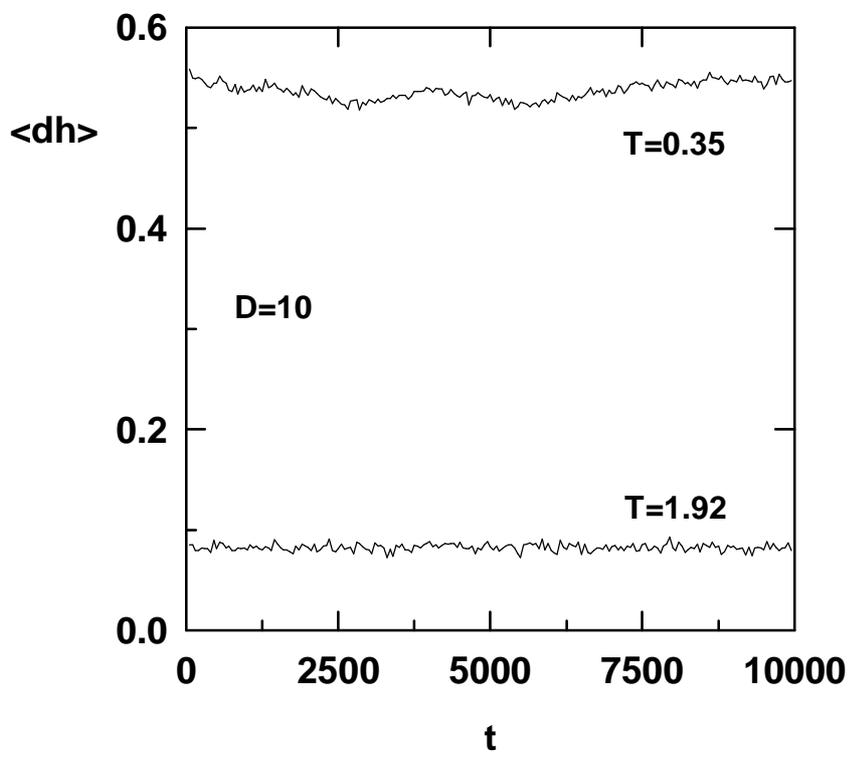
\begin{figure}
\label{fig6}
\caption{$\langle dh \rangle$ vs. $t$ for $D=10$, $dh(0)=1$ 
and for $T=0.35$ (low temperature phase) and $T=1.92$ (intermediate phase).}
\end{figure}

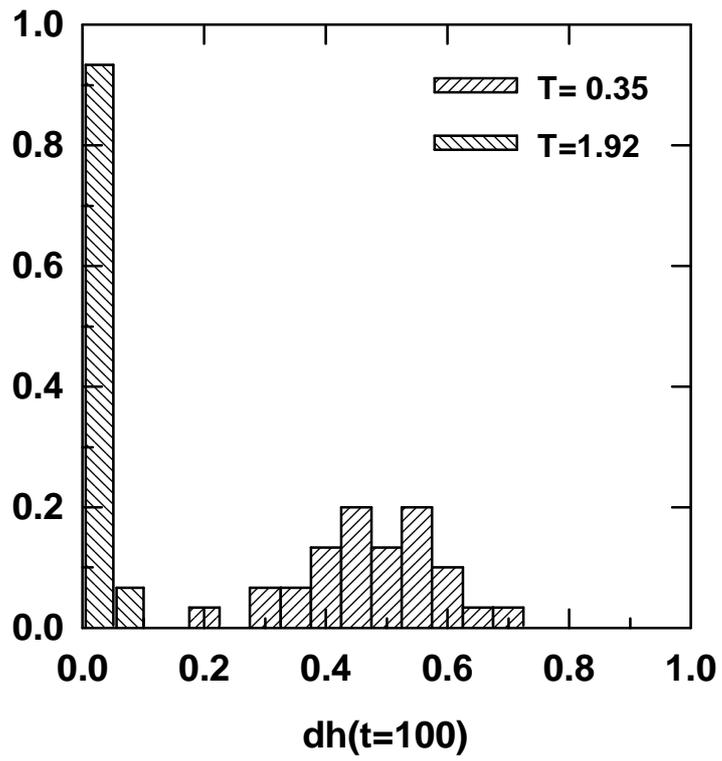
\begin{figure}
\label{fig7}
\caption{Histogram of Hamming distances at $t=100$ with $dh(0)=1.$
for $D=10$ and a) $T=0.35$ (in the low temperature phase) and 
b) $T=1.92$ (in the intermediate phase).}
\end{figure}

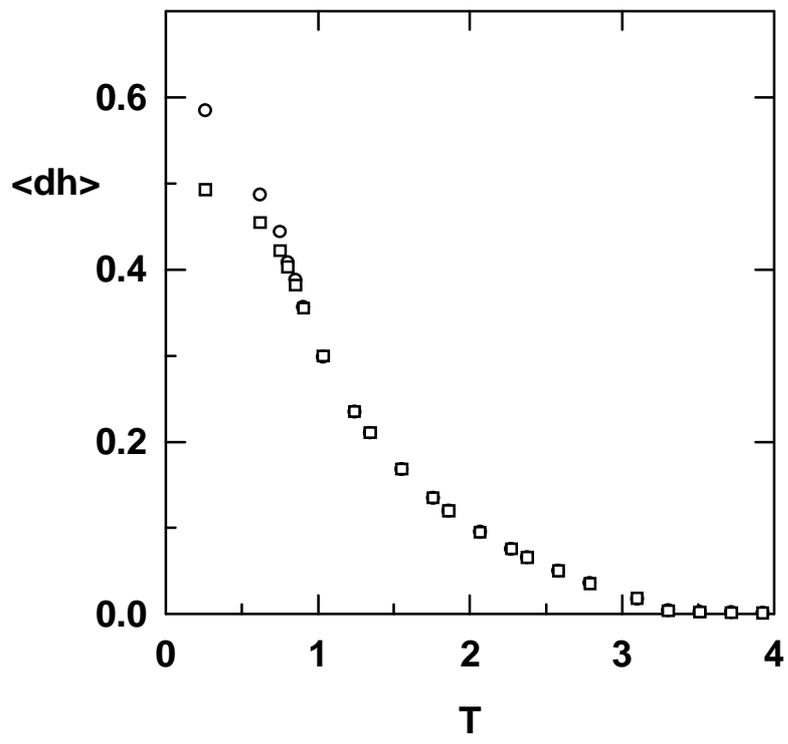
\begin{figure}
\label{fig8}
\caption{$\langle dh \rangle$ vs Temperature for $D=15$ and for two
different initial damages: $dh(0)=1$ (circles), $dh(0)=0.5$
(squares).}
\end{figure}

%%%%%%%%%%%%%%%%%%%%%%%%%%%%%%%%%%%%%%%%%%%%%%%%%%%%%%%%%%%%%%%%
%          Caption for tables
%%%%%%%%%%%%%%%%%%%%%%%%%%%%%%%%%%%%%%%%%%%%%%%%%%%%%%%%%%%%%%%%
%

\begin{table}
\label{tabla1}
\begin{center}
\begin{tabular}{|l|l|l|}
\hline
$D$ & $T_{1}^{D}$ & $T_{2}^{D}$ \\
\hline
$6$ & $0.65 \pm 0.04$  & $1.8 \pm 0.2$ \\
\hline
$8$ & $0.66 \pm 0.04$ & $1.7 \pm 0.1$ \\
\hline
$10$& $0.74 \pm 0.08$ & $2.0 \pm 0.1$ \\
\hline
$15$& $0.79 \pm 0.01$ & $3.25 \pm 0.05$\\
\hline
\end{tabular}
\end{center}
\end{table}

%%%%%%%%%%%%%%%%%%%%%%%%%%%%%%%%%%%%%%%%%%%%%%%%%%%%%%%%%%%%%%%%
%          References
%%%%%%%%%%%%%%%%%%%%%%%%%%%%%%%%%%%%%%%%%%%%%%%%%%%%%%%%%%%%%%%%
%


\begin{references}
% 1
\bibitem{Fischer} K.H. Fischer and J.A. Hertz, {\em Spin Glasses},
Cambridge University Press, Great Britain, 1993
% 2
\bibitem{Edwards} S.F. Edwards and P.W. Anderson, {\em J. Phys.} {\bf
F 5}, 965 (1975)
% 3
\bibitem{Sherrington} D. Sherrington and S. Kirkpatrick, {\em Phys.
Rev. Lett.} {\bf 35}, 1792 (1975)
% 4
\bibitem{Ogie} Ogielski A.T. {\em Phys. Rev. Lett. }{\bf 54}, 928 (1985)
% 5
\bibitem{Bray} A.J. Bray and M.A. Moore {\em J. Phys C:Solid State
Phys.}{\bf 17} L463, L613 (1984) \\
A.J. Bray {\em Comment Cond. Mat. Phys.} {\bf 14} 21 (1988)
% 6
\bibitem{pariru} Parisi G.,Ritort F., Rub\'{\i} J.M. (1991)
 {\em J.Phys.A:Math.Gen }{\bf24} 5307
 % 7
\bibitem{Cugliandolo} L. Cugliandolo and J. Kurchan, {\em J. Phys. A}
{\bf 27}, 5749 (1994)
% 8
\bibitem{Marinari} E. Marinari, G. Parisi and F. Ritort, {\em J. Phys.
A}{\bf 28}, 327 (1995)
% 9
\bibitem{Stariolo} D. Stariolo, preprint CONDMAT/9607132
% 7
\bibitem{derwei} Derrida B., Weisbuch G. (1987),
 {\it Europhys. Lett.} {\bf 4}, 657
% 8
\bibitem{Silva} L. da Silva, F.A. Tamarit and A.C.N. de Magalh\~aes,
{\em J. Phys. A} {\bf 30}, 2329 (1997)
% 9
\bibitem{arca} De Arcangelis L. (1990), in {\em Correlations and Connectivity},
 edited by Stanley H. E. and Ostrowsky N. (Kluver Acad. Publ.)

\end{references}
\end{document}